# DFT Investigation of pH-Driven Oxygen Vacancy Formation-Annihilation in CeO$_2$

Hongyang Ma,*,[a] Hangjuan Ren,[a] Zhao Liu,[b] Pramod Koshy,[a] Charles C. Sorrell,[a] and Judy N. Hart*,[a]

[a] School of Materials Science and Engineering, UNSW Sydney, NSW 2052, Australia

[b] Sino-French Institute of Nuclear Engineering and Technology, Sun Yat-sen University, Zhuhai 519082, China

*Corresponding Authors: hongyang.ma@unsw.edu.au, j.hart@unsw.edu.au

**Abstract**

There is considerable interest in the pH-dependent switchable biocatalytic properties of cerium oxide nanoparticles (CeNPs) in biomedicine, where these materials exhibit beneficial antioxidant activity against reactive oxygen species at neutral and basic physiological pH but cytotoxic prooxidant activity at acidic pathological pH. Oxygen vacancies play a key role in such biocatalytic activities. While the general characteristics of the role of oxygen vacancies are known, the mechanism of their action at the atomic scale under different pH conditions has yet to be elucidated. The present work applies density functional theory (DFT) calculations to interpret the pH-induced behavior of the stable {111} surface of $CeO_2$ at the atomic scale. Analysis of the surface-adsorbed media species reveals the critical role of pH on the reversibility of the $Ce^{3+} \leftrightarrow Ce^{4+}$ redox equilibria and the formation $\leftrightarrow$ annihilation of the oxygen vacancies. Under acidic conditions, this reversible switching is hindered owing to incomplete volumetric filling of the oxygen vacancies by the oxygen in the water molecules, hence effective retention of the oxygen vacancies, and consequent inhibition of redox biomimetic reactions. Under neutral and basic conditions, the capacity for this reversible switching is preserved due to complete filling of the oxygen vacancies by the $OH^-$ ions owing to their ready size accommodation, thereby retaining the capacity for performing redox biomimetic reactions cyclically.

**Introduction**

Cerium oxide nanoparticles (CeNPs) have attracted considerable attention and commonly are used for a variety of catalytic, photocatalytic, and biomedical applications[1-4]. A key attribute of CeNPs is the rapid and reversible switching ability of the oxidation states between $Ce^{4+}$ and $Ce^{3+}$, where $Ce^{3+}$ and the charge-compensating oxygen vacancies ($V_O^{\bullet\bullet}$) localize principally on the nanoparticle surface[5]. This allows CeNPs to alter their surface electronic configuration in response to different microenvironments with negligible restructuring of the fluorite crystal structure[6]. This reversibility, which is facilitated by the low redox potential of $Ce^{4+}$ and $Ce^{3+}$ switching[7], provides a multi-functional performance of CeNPs and confers the capacity for self-regeneration of the catalytic activity, which has allowed CeNPs to be employed in a variety of industrial applications, including catalysis[8, 9], UV sunscreen[10], and solar cells[11].

This regenerative ability of CeNPs also is responsible for their multi-catalytic biomimetic behavior[12], allowing CeNPs to exhibit: (1) antioxidative property by switching $Ce^{4+}$ to $Ce^{3+}$ to decompose reactive oxygen species (ROS) to harmless products ($H_2O$ and $O_2$), and (2) prooxidative property by reverse switching $Ce^{3+}$ to $Ce^{4+}$ to generate harmful ROS. This gives CeNPs a multi-functional performance capacity, including: (1) catalase (CAT) and phosphatase biomimetic reactivities, whereby $CeO_2$ acts as an antioxidant, and (2) superoxide dismutase (SOD), peroxidase, and oxidase biomimetic reactivities, whereby $CeO_2$ acts as a prooxidant[13]. Furthermore, unlike pharmacological agents or enzymes, which generally can provide only one active site per molecule, CeNPs can exhibit significantly higher numbers of active sites in the form of $V_O^{\bullet\bullet}$. The concentration of $V_O^{\bullet\bullet}$ can be modified flexibly through variations in processing and associated exposed crystallographic planes[14-16]. Consequently, CeNPs have emerged as important functional materials in biomedicine[17], particularly for drug delivery[18, 19], bioanalysis[20], and bioscaffolding[21].

Although CeNPs have shown tremendous potential in biomedical applications, the mechanisms associated with the factors that affect their multi-catalytic behavior are poorly understood. The variables that can affect the performance include pH conditions of the microenvironment[22], exposed crystallographic planes[1], associated ζ-potential[23], particle size[24], and impurities and dopants[25, 26]. Among these, the role of pH in their multi-catalytic behavior has been studied in some depth[27], since the typical pH conditions in physiological microenvironments are basic (pH: 7.2-7.4[28], 7.35-7.45[29, 30], 7.0-8.0[31]) but that in pathological microenvironments, such as tumor cells, generally are acidic (pH: 6.0-6.5[32], 6.4-7.0[33], 6.5-

6.9[34]). Consequently, CeNPs can provide pH-driven, switchable, and biocatalytic functionalities[35, 36]:

(1) **Physiological Conditions:** CeNPs provide antioxidant activity to protect healthy cells from oxidative stress damage by elimination of ROS, thereby promoting cell proliferation[37, 38].
(2) **Pathological Conditions:** CeNPs provide prooxidant activity to destroy diseased cells by production of ROS, thereby facilitating intracellular oxidation and inducing cell apoptosis[39].

Thus, the combination of antioxidative and prooxidative behavior confers on CeNPs the ability to regulate ROS elimination and production as a function of the pH in the biological environments, and hence offers a potential route for various applications including bioanalysis (e.g., detection of chronic inflammation[40]) and biomedicine (e.g., treatment of retinal damage[41] and cancer[35]).

Although some researchers have reported and developed $CeO_2$-based pH-responsive assemblies that can effectively protect healthy cells while killing diseased cells, the mechanism associated with this selective killing performance remains unclear and there are still some exceptions that bring into question the universality of the selective cytotoxicity of CeNPs, with examples being reported of induced normal cell apoptosis[42] and cancer cell protection[37, 38]. These exceptions limit the attractive potentialities of CeNPs to laboratory and research settings and limit the clinical applications; therefore, more investigations are required urgently to gain a greater understanding of the mechanisms associated with the pH-induced multi-catalytic behavior of CeNPs, which then may provide a real diagnostic and therapeutic medicinal tool against cancer and other diseases.

The importance of $V_O^{\bullet\bullet}$ to the unique performance of CeNPs has been demonstrated[43-45]; however, there do not appear to be any comprehensive studies that have revealed the specific relationship between the pH and the formation and state of $V_O^{\bullet\bullet}$. Previous research has focused primarily on the study of the formation of $V_O^{\bullet\bullet}$ on the surface or subsurface of CeNPs, the localization of the resulting two extra 4f electrons, and the interaction between oxygen species (e.g., $O_2$, superoxide, *etc.*) and the pristine low-index $CeO_2$ surfaces[46-48]. However, data associated with the influence of aqueous environments and pH conditions ($H_3O^+$ and $OH^-$ ions) on the surface chemistry of $CeO_2$ and the formation of $V_O^{\bullet\bullet}$ are scarce, and neglecting such influences results in missing key factors that may play critical roles in the unique performance of $CeO_2$.

The present work employs density functional theory (DFT) calculations to reveal, at the atomic scale, the origins of the pH-induced behavior of CeNPs. Systematic investigations were performed to shed light on the adsorption of media molecules, including water molecules ($H_2O$), hydronium ions ($H_3O^+$), and hydroxide ions ($OH^-$), on the most stable $CeO_2$ {111} surface, and the resultant influence on the formation and annihilation of surface $V_O^{\bullet\bullet}$. This work may pave the way toward a better understanding of the pH-induced multi-catalytic performance of CeNPs, which will be beneficial for the optimization of the material's performance and provide better control in applications, such as cancer treatment.

**Computational Methods**

Spin-polarized periodic density functional theory (DFT) calculations were employed to investigate the influence of pH on the formation of $V_O^{\bullet\bullet}$ on the $CeO_2$ {111} surface. The calculations were performed with the CRYSTAL17 code[49] using a B3LYP-like method[50, 51] to calculate the exchange-correlation energy. Different amounts of Hartree-Fock (HF) exchange energy mixed with the DFT exchange-correlation energy were tested. The results showed that 12% HF energy could most accurately reproduce the experimentally measured band gaps and lattice parameters of bulk $CeO_2$ (Table S1, Supporting Information). The van der Waals interactions were included in all calculations, with C6 coefficients chosen according to the Grimme D3 scheme[52]. An effective core pseudopotential for cerium[53], an 8-411(d1) basis set for oxygen[54], and a TZVP basis set for hydrogen[55] were used. The Coulomb and exchange series were truncated with threshold values of $10^{-7}$, $10^{-7}$, $10^{-7}$, $10^{-7}$, and $10^{-14}$. The convergence threshold on energy for the self-consistent-field (SCF) procedure was set to $1.0\times10^{-7}$ Hartree (Ha). For the geometry optimization, the convergence criteria for the root mean square (RMS) force and RMS atomic displacement values were set to $3 \times 10^{-4}$ Ha/Bohr and $1.2 \times 10^{-3}$ Bohr, respectively, while the convergence criteria for the maximal force and maximal atomic displacement were 1.5 times their respective RMS values.

Slabs exposing the {111} surface were cleaved from an optimized bulk structure. To yield convergent surface energies, a slab with a thickness of three O−Ce−O trilayers was used for all calculations, followed by geometry optimization of both atomic coordinates and cell parameters with atoms in the bottom O−Ce−O trilayer frozen at their bulk positions. In order to study the influence of pH on the formation of $V_O^{\bullet\bullet}$, a 3 × 3 supercell of the optimized slab was generated, providing a ∼ 12 × 12 Å² surface area. This supercell size is large enough to avoid interactions between periodic images of $V_O^{\bullet\bullet}$. The *k*-point meshes were sampled using 8 × 8 × 8 and 2 × 2 × 1 Monkhorst-Pack grids for bulk $CeO_2$ and the supercell models of the slab structures, respectively; convergence with respect to the number of *k*-points was checked (Figure S1, Supporting Information).

Ten media molecules were placed above the $CeO_2$ {111} surface to simulate different pH conditions. The molecules included were ten $H_2O$ for neutral pH, and nine $H_2O$ and one $H_3O^+$/$OH^-$ for acidic/basic pH conditions. A uniform background charge was added in order to neutralize the charge (*i.e.*, of the $H_3O^+$/$OH^-$ ion) in the simulation cell. Ten molecules were sufficient to create one dense molecular layer covering the $CeO_2$ surface in the 3 × 3 supercell. Additionally, a test was conducted with six $H_2O$ molecular layers (sixty $H_2O$ molecules); the results showed no significant difference in the optimized structural arrangement of the layer of media molecules closest to the $CeO_2$ surface compared with the single $H_2O$ molecular layer (Figure S2, Supporting Information). Therefore, since the present work focuses on the interaction between the media molecules and the $CeO_2$ surface, ten molecules (*i.e.*, one molecular layer) were used in all subsequent calculations.

To study the influence of molecular arrangement on the adsorption of media molecules on the $CeO_2$ {111} surface, several different initial geometries for each pH condition were tested. The adsorption energies of media molecules on the $CeO_2$ {111} surface, $\Delta E_{ads}$, were calculated by:

$$\Delta E_{ads} = E_{slab/media} - E_{slab} - n_1 E_{H_2O} - n_2 E_{H_3O^+} - n_3 E_{OH^-} \quad (1)$$

where $E_{slab/media}$ and $E_{slab}$ are the energies of the slab with and without the media molecules, respectively; $n_1$, $n_2$, and $n_3$ represent the number of $H_2O$ molecules, $H_3O^+$ ions, and $OH^-$ ions in the system, respectively. $E_{H_2O}$, $E_{H_3O^+}$, and $E_{OH^-}$ are the energies of isolated $H_2O$ molecule, $H_3O^+$, and $OH^-$ ions, respectively.

After testing different geometries of the media molecules, the two most stable geometries for each pH condition were used to investigate the influence of pH conditions on the formation of $V_O^{\bullet\bullet}$. One surface oxygen was removed from the optimized $CeO_2$ {111} 3 × 3 surface supercell and the simulation cell was re-optimized. The formation energies of $V_O^{\bullet\bullet}$, $\Delta E_{V_O^{\bullet\bullet}}$, were calculated by:

$$\Delta E_{V_O^{\bullet\bullet}} = E_{slab/V_O^{\bullet\bullet}} - E_{slab} + \frac{1}{2} E_{O_2} \quad (2)$$

where $E_{slab/V_O^{\bullet\bullet}}$ and $E_{slab}$ are the energies of the defective and non-defective slabs under the same pH conditions, respectively, and $E_{O_2}$ is the energy of an isolated gas-phase O₂ molecule. For comparison, the formation of a $V_O^{\bullet\bullet}$ in the pristine CeO₂ {111} surface also was tested in the same way, *i.e.*, by removing one surface oxygen in the absence of adsorbed media molecules.

**Results and Discussion**

**Defective CeO₂ {111} Surface with No Media Molecules**

Three different geometries for the pristine CeO₂ {111} surface containing one $V_O^{\bullet\bullet}$, with no media molecules present, were tested. Creation of a $V_O^{\bullet\bullet}$ will leave two excess electrons in the lattice and thus result in the creation of two $Ce^{3+}$. It should be noted that, in general, it is not energetically favorable for the excess two electrons to localize on the Ce in the nearest neighbor positions to the $V_O^{\bullet\bullet}$ [56-58]; however, determining the correct positions of the resultant $Ce^{3+}$ is not straightforward and the correct positions may not be found spontaneously in the DFT calculation owing to the presence of local minima. Therefore, different locations of the two $Ce^{3+}$ relative to the location of $V_O^{\bullet\bullet}$ (*i.e.*, different distributions of the two excess electrons driven from the formation of $V_O^{\bullet\bullet}$) were tested, including the two electrons localized on: (1) two of the Ce that are the nearest neighbors to the $V_O^{\bullet\bullet}$ (labeled as NN-NN), (2) one Ce that is a nearest neighbor and another Ce that is a next-nearest neighbor to the $V_O^{\bullet\bullet}$ (labeled NN-NNN), and (3) two Ce that are next-nearest neighbors to the $V_O^{\bullet\bullet}$ (labeled NNN-NNN). To ensure localization of the extra electrons on certain Ce, all the oxygen immediately surrounding that Ce were pulled away manually by ~ 0.1 Å (*i.e.*, the Ce–O bonds were stretched by ~ 0.1 Å) before proceeding with the geometry optimization[57, 59] (Table S2 and Figure S3).

The results indicate that the case in which the two excess electrons localize on the NN-NN sites is the least favorable (vacancy formation energy of 4.57 eV), consistent with previous reports[46, 60]. The initial geometry with NNN-NNN sites optimized spontaneously to the same geometry as the NN-NNN case, with one electron moving from NNN site to NN site; therefore, the NN-NNN and NNN-NNN cases give almost identical formation energies of $V_O^{\bullet\bullet}$ (4.46 eV and 4.47 eV, respectively). In contrast, in previous studies, it was reported that both of the excess electrons locating on NNN sites is the most favorable[46, 58]. The different result obtained in the present work is caused presumably by the difference in the size of supercells used for the calculations. Compared with the previous work, the present work uses a smaller 3 × 3 supercell (~ 12 × 12 Å² surface area), which results in two NNN Ce localizing adjacent to each other across the periodic cell boundary, and this results in some interaction between the two $Ce^{3+}$. However, a 3 × 3 supercell can provide a 1/9 (11%) surface concentration of $V_O^{\bullet\bullet}$, which corresponds to the typical vacancy concentration in experimental work in nanooctahedra $CeO_2$ (exposing {111} surfaces)[61], thereby, it is an appropriate supercell size for the current study.

To further evaluate the most favorable location of $V_O^{\bullet\bullet}$, a subsurface $V_O^{\bullet\bullet}$ ($2^{nd}$-layer) was generated; this was found to give a lower formation energy of $V_O^{\bullet\bullet}$ (4.30 eV), indicating that on a pristine $CeO_2$ {111} surface (with no media molecules present), the $2^{nd}$-layer $V_O^{\bullet\bullet}$ is more favorable than the $1^{st}$-layer $V_O^{\bullet\bullet}$, consistent with previous reports[60]. However, as discussed in a later section, when the influence of media molecules present on the surface is considered, the formation of $1^{st}$-layer $V_O^{\bullet\bullet}$ is found to be more favorable.

**Non-Defective $CeO_2$ {111} Surface with Media Molecules**

In order to study the influence of different arrangements of media molecules on the $CeO_2$ {111} surface, three different initial geometries with randomly arranged media molecules were tested for each of three different pH conditions. The calculated adsorption energies of media molecules are shown in Table 1 and the optimized geometries are shown in Figure S4.

**Table 1.** Adsorption energies of media molecules on non-defective $CeO_2$ {111} surface under different pH conditions

|  | Adsorption Energy (kcal/mol) | | |
| --- | --- | --- | --- |
|  | Geometry 1 | Geometry 2 | Geometry 3 |
| Neutral | -219.60 | -220.61 | -223.98 |
| Acidic | -385.30 | -388.11 | -389.04 |
| Basic | -391.71 | -392.02 | -392.74 |

*Neutral:* Similar adsorption energies with a variation of less than 5 kcal/mol between the three geometries (Table 1 and Figure S4) indicate that the arrangement of the media molecules has negligible influence under neutral conditions. To further investigate the influence of the number of layers of media molecules on the interaction between the $H_2O$ molecules and the $CeO_2$ {111} surface, five additional molecular layers of $H_2O$ were added on top of the optimized monolayer and the structure was re-optimized. As shown in Figure S2, the arrangement of the layer of $H_2O$ molecules closest to the surface remains similar to that for the case of the single molecular layer. Therefore, it seems that the number of layers of media molecules has no significant influence on the interaction between the media and the $CeO_2$ surface.

*Acidic and basic:* Both acidic and basic pH conditions show negligible differences in the adsorption energies with various arrangements of the media molecules on the $CeO_2$ {111} surface, consistent with the result for neutral conditions. The difference between the adsorption energies for acidic and basic conditions is negligible, but the adsorption of media molecules under acidic and basic conditions is significantly more energetically favorable than under neutral conditions (Table 1). This is attributed to the stabilization of $O_{surface}$ and $Ce_{surface}$ by media molecules under acidic and basic conditions. For the $CeO_2$ {111} surface, the coordination numbers of $O_{surface}$ and $Ce_{surface}$ are three and seven, respectively, while for bulk $CeO_2$, these numbers are four and eight, respectively; thus, the surface atoms are under-coordinated. Under neutral conditions, while some $Ce_{surface}$–$O_{water}$ bonds do form to partially recover the coordination numbers of $Ce_{surface}$ and stabilize $Ce_{surface}$, these bonds are quite weak (~2.7 Å). In contrast, under acidic conditions, in addition to the $Ce_{surface}$–$O_{water}$ bonds, a strong covalent bond forms between $O_{surface}$ and proton that dissociates from the $H_3O^+$ (Figure S4); this bond, which has a length of 0.97 Å, stabilizes $O_{surface}$, recovering the coordination number of $O_{surface}$ back to four. Under basic conditions, strong $Ce_{surface}$–$OH^-$ bonds (~2.4 Å) form, recovering the coordination number of $Ce_{surface}$ back to eight. Therefore, the adsorption of media molecules on the $CeO_2$ {111} surface under non-neutral conditions is more favorable than under neutral conditions.

Additionally, for non-neutral conditions, initial geometries with the $H_3O^+$ and $OH^-$ ions positioned at different distances from the surface were considered. For acidic conditions, regardless of the initial distance between the $CeO_2$ surface and the $H_3O^+$ ion, the proton always dissociates from the $H_3O^+$ ion and bonds to surface oxygen. When the $H_3O^+$ ion initially is far from the surface, this occurs spontaneously *via* a concerted motion of the proton through the $H_2O$ molecule network (Figure 1(a-d)). Similarly, in the case of basic conditions, the $OH^-$ ion will bond to surface cerium, regardless of the initial distance between the $CeO_2$ surface and the $OH^-$ ion, through a concerted $OH^-$-$H_2O$ interaction (Figure 1(e-h)). These phenomena indicate that the delivery of $H^+$ and $OH^-$ ions through the $H_2O$ molecule network to the $CeO_2$ surface is a rapid, exothermic process that does not require any activation energy. Therefore, all the following studies of defective $CeO_2$ surfaces under acidic and basic conditions are based on initial geometries with the $H_3O^+$ and $OH^-$ ions located close to the surface.

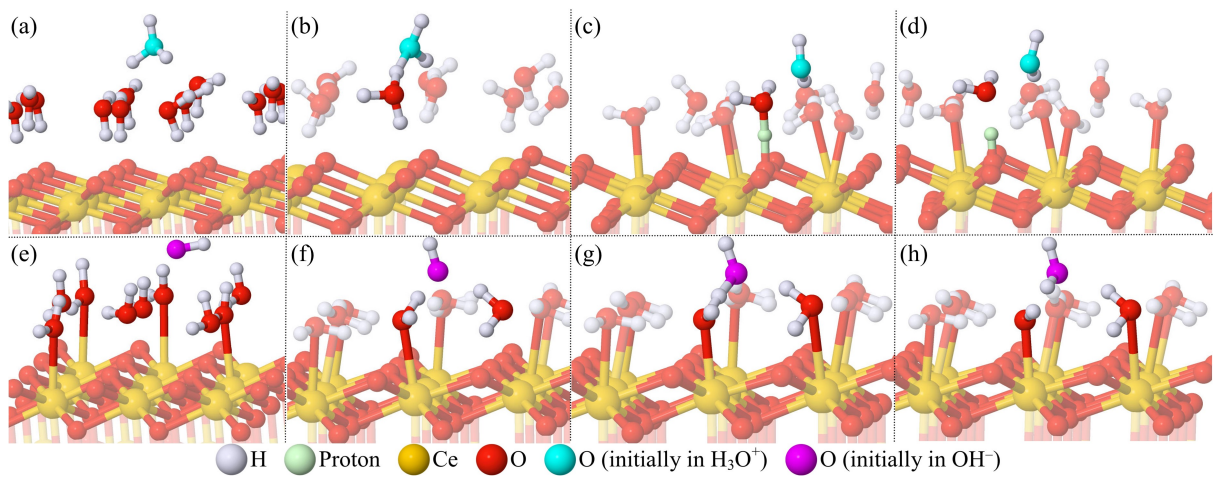

**Figure 1.** Sequential structural changes between initial and final geometries when $H_3O^+$ and $OH^-$ ions are placed initially far from $CeO_2$ {111} surface: (a-d) acidic conditions, highlighting trajectory of $H_3O^+$ ion and (e-h) basic conditions, highlighting trajectory of $OH^-$ ion

**Defective CeO₂ {111} Surface with Media Molecules – Selection of Defect Site**

As discussed above, since the NN-NNN arrangement of the $Ce^{3+}$ is most favorable, all the following calculations of defective systems used the NN-NNN arrangement as the initial geometry. Although the formation of the 2$^{nd}$-layer $V_O^{\bullet\bullet}$ was found to be energetically more favorable than the 1$^{st}$-layer $V_O^{\bullet\bullet}$ on a pristine $CeO_2$ {111} surface, studies of defective surfaces in the presence of media used an initial geometry with 1$^{st}$-layer $V_O^{\bullet\bullet}$ for the following reasons. Firstly, at room temperature or human body temperature, the diffusion of $V_O^{\bullet\bullet}$ between the 1$^{st}$-layer and 2$^{nd}$-layer is considered as a rapid and thermodynamically favorable process based on the similar formation energies of the two $V_O^{\bullet\bullet}$ locations (4.46 eV *vs.* 4.30 eV), entropy considerations, and the high $V_O^{\bullet\bullet}$ mobility[62]. Furthermore, the investigation of 1$^{st}$-layer $V_O^{\bullet\bullet}$ is more important for understanding catalytic reactions, since catalytic reactions generally take place on the surface.

An additional aspect to take into consideration is the influence of media molecules on the formation of $V_O^{\bullet\bullet}$, which has not been considered in most previous studies. The formation of the 1$^{st}$-layer $V_O^{\bullet\bullet}$ may be facilitated significantly by the presence of media molecules, due to the interactions between media molecules and the $CeO_2$ surface, which may stabilize the $V_O^{\bullet\bullet}$[63]. Therefore, to test the influence of media molecules on the formation of $V_O^{\bullet\bullet}$, two structures were considered – both had the same initial arrangements of $H_2O$ molecules (at neutral pH), while one had a 2$^{nd}$-layer $V_O^{\bullet\bullet}$ and the other had a 1$^{st}$-layer $V_O^{\bullet\bullet}$. The formation energy of the 2$^{nd}$-layer $V_O^{\bullet\bullet}$ in the presence of media molecules is identical to that for the pristine $CeO_2$ surface (4.30 eV in both cases), indicating the formation of the 2$^{nd}$-layer $V_O^{\bullet\bullet}$ is hardly influenced by the media molecules. In contrast, a significant enhancement is seen in the formation of the 1$^{st}$-layer $V_O^{\bullet\bullet}$, with the formation energy decreasing from 4.46 eV (no media) to 3.87 eV (with media); as discussed later, this large decrease in the formation energy of $V_O^{\bullet\bullet}$ can be attributed to the accommodation of media molecules in the $V_O^{\bullet\bullet}$, with the formation of $Ce_{surface}$–$O_{water}$ bonds, hence stabilizing the $V_O^{\bullet\bullet}$. Therefore, in the presence of media molecules, the formation of the 1$^{st}$-layer $V_O^{\bullet\bullet}$ is significantly more favorable than the formation of the 2$^{nd}$-layer $V_O^{\bullet\bullet}$.

Further evidence of the unfavorability of the 2$^{nd}$-layer $V_O^{\bullet\bullet}$ in the presence of media is that, as shown in Figure 2, there is an obvious trend that one surface oxygen near the subsurface $V_O^{\bullet\bullet}$ tends to move from the surface towards the subsurface during the geometry optimization (*i.e.*, the $V_O^{\bullet\bullet}$ tends to move from the 2$^{nd}$-layer towards the 1$^{st}$-layer). However, spontaneous movement of this $V_O^{\bullet\bullet}$ all the way to the 1$^{st}$-layer site is not seen. This may be due to the existence of an energy barrier that prevents the movement of surface oxygen through the closely compacted surface geometry of the CeO$_2$ {111} surface. However, at room temperature or human body temperature, it is highly possible that surface oxygen could overcome this energy barrier and complete the motion of $V_O^{\bullet\bullet}$ from the subsurface to surface, since the 1$^{st}$-layer $V_O^{\bullet\bullet}$ is energetically more stable than the 2$^{nd}$-layer $V_O^{\bullet\bullet}$ in the presence of media molecules. Therefore, in the following work, only 1$^{st}$-layer $V_O^{\bullet\bullet}$ are considered.

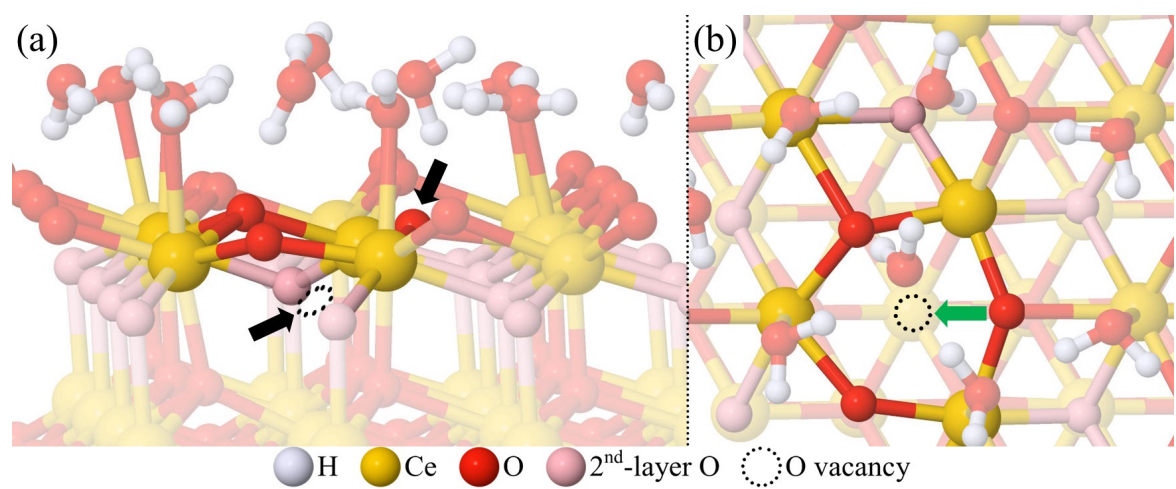

**Figure 2.** Optimized geometry for an oxygen vacancy ($V_O^{\bullet\bullet}$) created in subsurface (2$^{nd}$-layer) of CeO$_2$ {111} surface; black arrows highlight $V_O^{\bullet\bullet}$ and 1$^{st}$-layer oxygen that moves towards it during geometry optimization, and green arrow indicates direction of motion of surface oxygen; (a) side view and (b) top view

**Defective CeO$_2$ {111} Surface with Media Molecules – Effect of pH on Vacancy Formation**

In contrast to the results obtained for non-defective systems, a significant influence of the arrangement of media molecules on energetics is observed for defective surfaces; hence, the formation energies of $V_O^{\bullet\bullet}$ vary for different initial geometries of the media molecules. However, the overall trend is clear: the formation of $V_O^{\bullet\bullet}$ is generally most favorable under basic conditions, while it is least favorable under neutral conditions (Table 2). The reasons for this are discussed below.

**Table 2.** Formation energies of oxygen vacancies ($V_O^{\bullet\bullet}$) on CeO$_2$ {111} surface under different pH conditions; formation energies of H$^+$-far-vacancy (HFV) for acidic conditions and OH$^-$-far-vacancy (OHFV) for basic conditions are highlighted in **bold**

|  | Formation Energy of $V_O^{\bullet\bullet}$ (eV) | | | |
| --- | --- | --- | --- | --- |
|  | Geometry 1 | Geometry 2 | Geometry 3 | Geometry 4 |
| Neutral | 3.89 | 3.87 | 3.06[a] | -- |
| Acidic | **3.32** | **2.88** | 3.46 | 3.38 |
| Basic | **3.54** | 2.37[b] | 3.03 | 2.70 |

[a] In this mode, H$_2$O molecules spontaneously dissociated.
[b] In this mode, OHMV spontaneously optimized to OHNV.

*Neutral:* Three geometries with different arrangements of the media molecules were tested to investigate the formation of $V_O^{\bullet\bullet}$ under neutral conditions (Figure 3). Interestingly, in one case (Geometry 3 in Table 2 and Figure 3), spontaneous dissociation of one H$_2$O molecule near the $V_O^{\bullet\bullet}$ is observed, leaving one OH$^-$ ion that forming two Ce$_{surface}$–O bonds near the $V_O^{\bullet\bullet}$, and one H$^+$ ion that forming one O$_{surface}$–H bond adjacent to the $V_O^{\bullet\bullet}$. This indicates that the surface $V_O^{\bullet\bullet}$ can act as an active site for H$_2$O dissociation and accommodate the dissociated OH$^-$ ion.

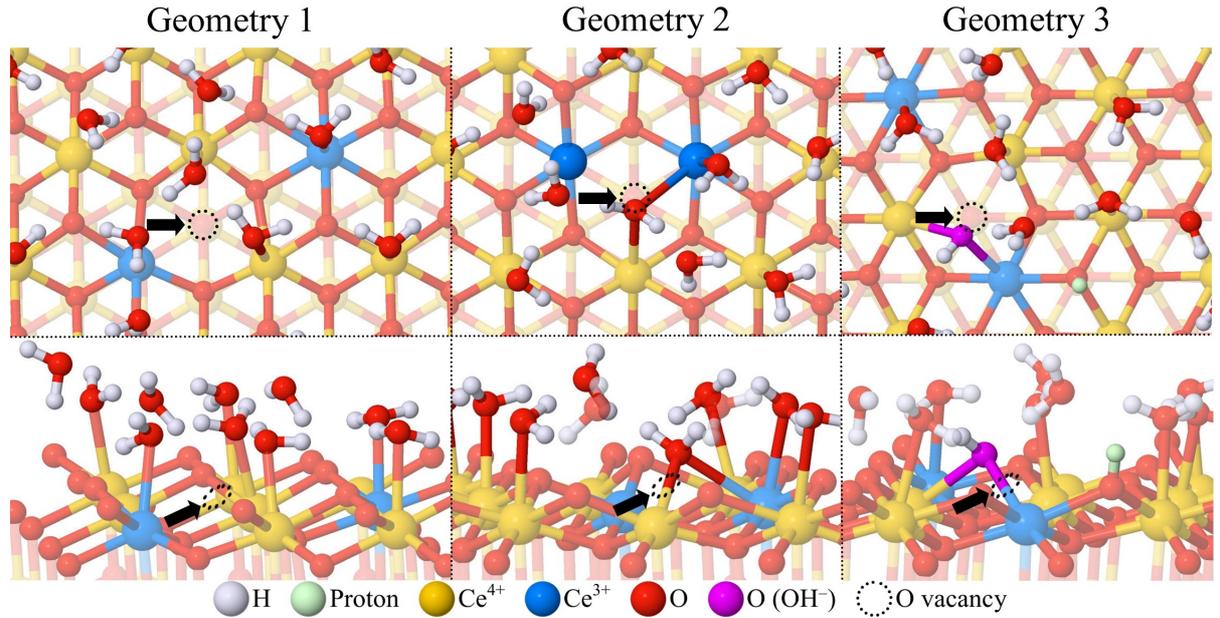

**Figure 3.** Optimized geometries for media molecules adsorbed on defective CeO$_2$ {111} surface (*i.e.*, containing an oxygen vacancy ($V_O^{\bullet\bullet}$)) under neutral conditions (top views in upper row; corresponding side views in lower row); black arrows indicate locations of $V_O^{\bullet\bullet}$

In the other two geometries considered, no dissociation of H₂O molecules is seen. These two geometries give similar formation energies of $V_O^{\bullet\bullet}$ with only 0.02 eV difference (Table 2). It should be noted that, although both cases give similar formation energies of $V_O^{\bullet\bullet}$, the optimized geometries are slightly different. In one case, one H₂O molecule near the $V_O^{\bullet\bullet}$ is accommodated partially in the $V_O^{\bullet\bullet}$ (Geometry 2 in Figure 3), forming two Ce$_{surface}$–O$_{water}$ bonds (2.65 Å and 2.89 Å) with the Ce adjacent to the $V_O^{\bullet\bullet}$. However, no such interaction with the $V_O^{\bullet\bullet}$ is seen in the other case (Geometry 1 in Figure 3), for which the minimal Ce$_{surface}$–O$_{water}$ bond length found anywhere on the surface is 2.70 Å; this may be due to the closely packed arrangement of three H₂O molecules around the $V_O^{\bullet\bullet}$, which hinders any one of the molecules from filling the $V_O^{\bullet\bullet}$. However, the similar formation energies of $V_O^{\bullet\bullet}$ in these two cases indicate that, under neutral conditions, any differences in the strength of interactions between H₂O molecules and the CeO₂ surface associated with such minor differences in geometry are small and do not have a large effect on the surface stability.

The geometry for which spontaneous H₂O dissociation occurred (Geometry 3) has a significantly smaller formation energy of $V_O^{\bullet\bullet}$ than the other two geometries for which there is no H₂O dissociation. This indicates that spontaneous dissociation of H₂O molecules on a defective CeO₂ surface and the resultant accommodation of OH⁻ ions in the $V_O^{\bullet\bullet}$ can occur and is energetically favorable under neutral conditions, but the dissociation process may be influenced to a certain extent by the arrangement of media molecules and only certain arrangements allow this dissociation to occur with no energy barrier.

Finally, it should be noted that the positions of the Ce³⁺ in two of the three tested geometries are consistent with that discussed above for pristine defective CeO₂ surfaces (*i.e.*, the Ce³⁺ localized spontaneously at the NN-NNN sites), while for Geometry 2, the NN-NN arrangement is observed. However, no firm conclusion about which sites are most favorable for the Ce³⁺ in the presence of media can be drawn from this work, since not all of the possible arrangements were tested.

*Acidic:* Four different media adsorption modes were considered on a defective surface under acidic conditions. To investigate the effect of the location of the H⁺ ion on the formation of $V_O^{\bullet\bullet}$, two situations were considered: (1) the dissociated proton from H₃O⁺ was placed manually on an O$_{surface}$ that neighbors one of Ce adjacent to the $V_O^{\bullet\bullet}$, labeled as H⁺-near-vacancy (HNV) and (2) the dissociated proton from H₃O⁺ was placed manually on an O$_{surface}$ that is far from Ce adjacent to the $V_O^{\bullet\bullet}$, labeled as H⁺-far-vacancy (HFV) (Figure S5). In each case, two different initial geometries for the media molecules were tested.

In contrast to the results for all the non-defective surfaces and the neutral pH cases for the defective surface without $H_2O$ dissociation, the arrangement of media molecules under acidic conditions has a significant effect on the formation of $V_O^{\bullet\bullet}$ (Table 2). Compared with neutral conditions, a clear reduction in the formation energy of $V_O^{\bullet\bullet}$ under acidic conditions indicates that the incorporation of protons ($H^+$) can enhance the formation of $V_O^{\bullet\bullet}$, consistent with previously reported experimental work[64]. Additionally, similar to neutral conditions discussed above, the excess electrons left by the removal of oxygen tend to localize spontaneously in the NN-NNN sites, although in one case (HFV-Geometry 1), the NNN-NNN arrangement is found.

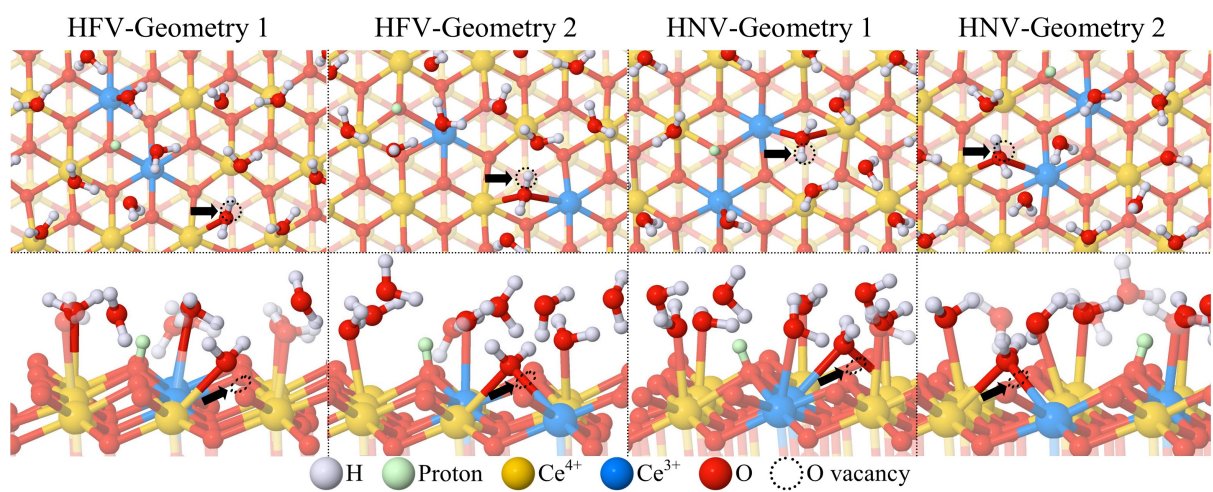

**Figure 4.** Optimized geometries for media molecules adsorbed on defective $CeO_2$ {111} surface (*i.e.*, containing an oxygen vacancy ($V_O^{\bullet\bullet}$)) under acidic conditions (top views in upper row; corresponding side views in lower row); black arrows indicate locations of $V_O^{\bullet\bullet}$

As for the non-defective surface, in all cases of defective surfaces at acidic pH, a proton dissociates from the $H_3O^+$ ion and bonds to surface oxygen. As seen under neutral conditions, under acidic conditions it is observed that the media $H_2O$ molecules tend to be accommodated partially in the $V_O^{\bullet\bullet}$ to stabilize the $V_O^{\bullet\bullet}$, forming one or two bonds with Ce adjacent to the $V_O^{\bullet\bullet}$ (Figure 4). However, in contrast to neutral conditions, in none of the four geometries tested under acidic conditions does spontaneous dissociation of $H_2O$ molecules occur. This smaller possibility of spontaneous dissociation may be because dissociation of an $H_2O$ molecule creates an $H^+$ ion that would be likely to form a covalent bond with surface oxygen; however, in the acidic cases, there are already $H^+$ ions bonded to surface oxygen (*i.e.*, from protons that were initially in $H_3O^+$ species), and a further increase in the surface coverage of $H^+$ ions may be unfavorable. Therefore, under acidic conditions, $V_O^{\bullet\bullet}$ would be more likely to be filled only partially by $H_2O$ molecules rather than being filled by $OH^-$ ions from $H_2O$ dissociation.

*Basic:* Compared with the other two pH conditions, the arrangement of surface media molecules has a more significant influence on the formation energy of $V_O^{\bullet\bullet}$ under basic conditions (Table 2). Similar to the acidic cases, three types of initial geometries with different distances between OH⁻ ion and $V_O^{\bullet\bullet}$ were tested, and are labeled as OH⁻-far-vacancy (OHFV, OH⁻ ion adsorbs on the Ce far from the $V_O^{\bullet\bullet}$), OH⁻-middle-vacancy (OHMV, OH⁻ ion adsorbs on NNN Ce), and OH⁻-near-vacancy (OHNV, OH⁻ ion adsorbs on NN Ce) (Figure S5 and Figure 5). As discussed below, the OHNV case is found to be most energetically favorable, so one additional OHNV case with a different initial arrangement of the media molecules was tested.

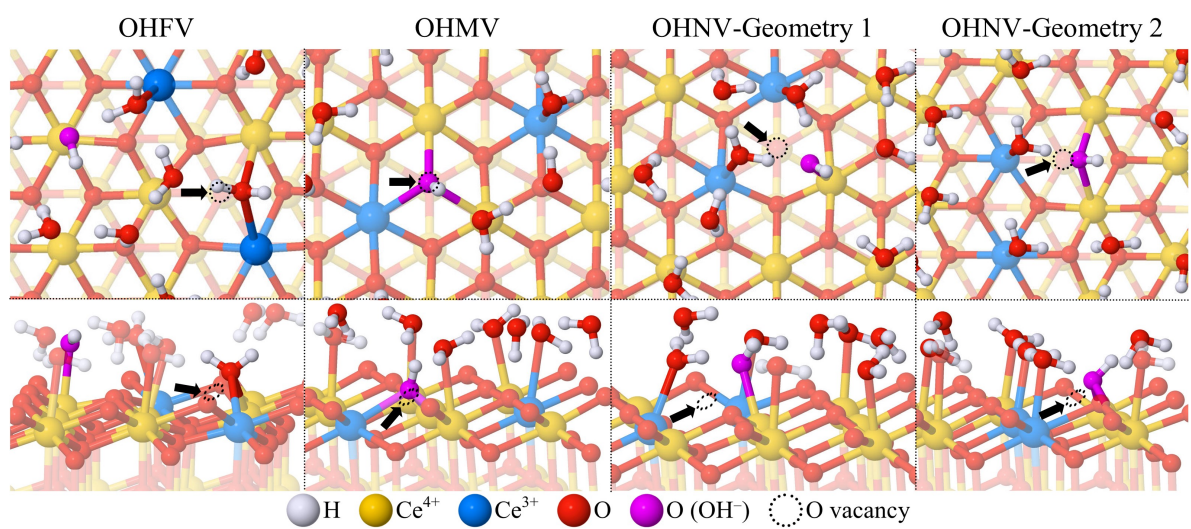

**Figure 5.** Optimized geometries for media molecules adsorbed on defective CeO₂ {111} surface (*i.e.*, containing an oxygen vacancy ($V_O^{\bullet\bullet}$)) under basic conditions (top views in upper row; corresponding side views in lower row); black arrows indicate locations of $V_O^{\bullet\bullet}$

In the OHMV case, the OH⁻ ion, which originally is slightly removed from the $V_O^{\bullet\bullet}$ (though not as far removed as in the OHFV case), spontaneously transfers from its original location to the location of the $V_O^{\bullet\bullet}$ *via* a concerted interaction between the OH⁻ ion and the H₂O molecule that originally is closest to the $V_O^{\bullet\bullet}$ which is dissociated during the process (Figure 6). In the optimized geometry, the OH⁻ ion created by the H₂O dissociation fills the $V_O^{\bullet\bullet}$, forming three Ce$_{surface}$–O$_{OH^-}$ bonds with Ce$_{surface}$ adjacent to the $V_O^{\bullet\bullet}$ and thus recovering the coordination number of Ce back to seven, as in the non-defective CeO₂ {111} surface. Thus, as seen for neutral conditions, $V_O^{\bullet\bullet}$ can act as centers for H₂O dissociation, and under basic conditions, OH⁻ ions close to $V_O^{\bullet\bullet}$ can facilitate this process.

In one of the OHNV cases (OHNV-Geometry 1 in Figure 5 and Geometry 3 in Table 2), both the OH⁻ ion and the H₂O molecules are excluded from the $V_O^{\bullet\bullet}$ due to the compact arrangement of H$_2$O molecules around the $V_O^{\bullet\bullet}$, while the other OHNV case (OHNV-Geometry 2 in Figure 5 and Geometry 4 in Table 2) is optimized to a similar structure as the OHMV case, with the OH$^-$ ion filling the $V_O^{\bullet\bullet}$; in this case, only two Ce$_{surface}$–O$_{OH^-}$ bonds are formed. In the OHFV case, no spontaneous delivery of the OH$^-$ ion to $V_O^{\bullet\bullet}$ is seen owing to the long distance and hence weak interaction between the OH$^-$ ion and $V_O^{\bullet\bullet}$. Instead, the OH$^-$ ion bonds with Ce$_{surface}$ away from the $V_O^{\bullet\bullet}$. Similar to the neutral and acidic conditions, the $V_O^{\bullet\bullet}$ is filled partially by an H$_2$O molecule in the OHFV case.

Under basic conditions, the smallest formation energy for $V_O^{\bullet\bullet}$ is found in the OHMV case, as it is in the case for neutral conditions in which an H$_2$O molecule spontaneously dissociates and $V_O^{\bullet\bullet}$ is filled with the OH$^-$ ion. Despite the similar geometry, the formation energy of $V_O^{\bullet\bullet}$ is slightly higher for OHNV-Geometry 2 than OHMV, suggesting that the interaction between the CeO$_2$ surface and the media molecules has a significant influence on the stability of the defective CeO$_2$ surface under basic conditions. In contrast, the OHFV case gives a noticeably higher formation energy of $V_O^{\bullet\bullet}$ than the OHMV and OHNV cases, indicating that $V_O^{\bullet\bullet}$ that is filled only partially by the H$_2$O molecule results in less stable surface chemistry. Overall, the results indicate that filling the $V_O^{\bullet\bullet}$ with OH$^-$ ions is more energetically favorable than leaving the $V_O^{\bullet\bullet}$ unfilled or partially filled by an H$_2$O molecule. This conclusion can be supported by the fact that H$_2$O molecules can, at best, fill only partially the $V_O^{\bullet\bullet}$ and form a maximum of two bonds with Ce$_{surface}$ (with lengths in the range 2.7-2.9 Å), owing to their large molecular size and two-fold coordinated oxygen, while OH$^-$ ions can fit perfectly in the $V_O^{\bullet\bullet}$ and form two to three bonds with Ce$_{surface}$ (2.4-2.6 Å).

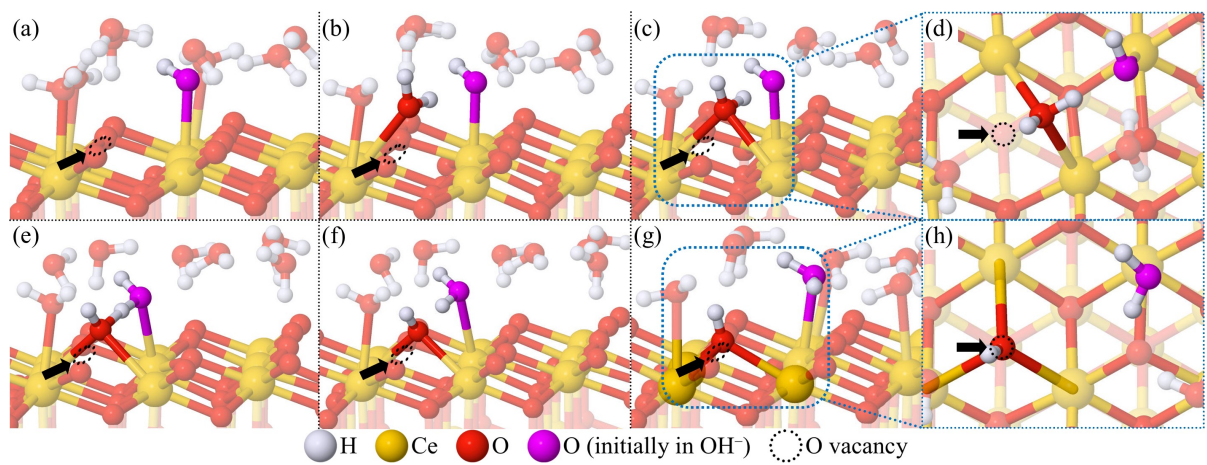

**Figure 6.** Sequential structural changes during optimization (a-c, e-g) of OHMV case on defective CeO$_2$ {111} surface (*i.e.*, containing one oxygen vacancy ($V_O^{\bullet\bullet}$)) under basic pH conditions; (d) and (h) are enlarged top views of enclosed regions in (c) and (g), respectively; black arrows indicate locations of $V_O^{\bullet\bullet}$

Similar to neutral and acidic conditions, the two excess electrons left by the removal of the oxygen tend to localize at the NN-NNN sites, with one exception (OHNV-Geometry 1) in which an NN-NN arrangement is seen. It should be noted that, although the $V_O^{\bullet\bullet}$ is annihilated by the OH$^-$ ion, recovering the CeO$_2$ surface to a non-defective-like structure and restoring the surface atom coordination numbers to their values in the non-defective surface, the Ce$^{3+}$ are not reset to Ce$^{4+}$ and still exist in the lattice.

**Mechanisms Associated with the pH-Controlled Biocatalytic Performance of CeNPs**

Differential inhibition of biomimetic reactions under different pH conditions has been proposed to play a critical role in the pH-controlled biocatalytic performance of CeNPs[27]. The pH-dependence of the reversibility between Ce$^{3+}$ and Ce$^{4+}$ is likely to be a key factor in the unique biocatalytic performance of CeO$_2$, since the switching between Ce$^{4+}$ and Ce$^{3+}$ states is the precondition for performing redox biomimetic reactions cyclically:

$$\text{Oxidation (Ox.):} \quad {}^{\bullet}O_2^- + 2H^+ + Ce^{3+} \rightarrow Ce^{4+} + H_2O_2 \qquad (3)$$
$$\text{Reduction (Re.):} \quad H_2O_2 + 2OH^- + 2Ce^{4+} \rightarrow 2Ce^{3+} + O_2 + 2H_2O \qquad (4)$$

The results found in this work demonstrate a significant difference in the surface chemistry of CeO$_2$ between neutral, basic, and acidic conditions. More specifically, the reversibility of the CeO$_2$ surface state is influenced considerably by pH. Under neutral and basic conditions (Figure 7), any existing $V_O^{\bullet\bullet}$ can be annihilated due to complete filling by OH$^-$ ions and the surface structure can be recovered to a state of non-defective-like CeO$_2$ {111}, leaving two extra electrons (*i.e.*, 2 Ce$^{3+}$) in the lattice (Step 1). Then, with an oxidation biomimetic reaction, the Ce$^{3+}$ are shifted to Ce$^{4+}$, thus fully recovering the surface structure to a non-defective CeO$_2$ {111} (Step 2). Subsequently, a reduction biomimetic reaction occurs, switching Ce$^{4+}$ back to Ce$^{3+}$ (Step 3). In this process, $V_O^{\bullet\bullet}$ are reformed, such that the CeO$_2$ surface is recovered to its original state (*i.e.*, defective CeO$_2$ {111}). Therefore, owing to the regenerability of $V_O^{\bullet\bullet}$ and the fully reversible switching between Ce$^{3+}$ and Ce$^{4+}$ under neutral and basic conditions, CeO$_2$ can perform redox biomimetic reactions cyclically.

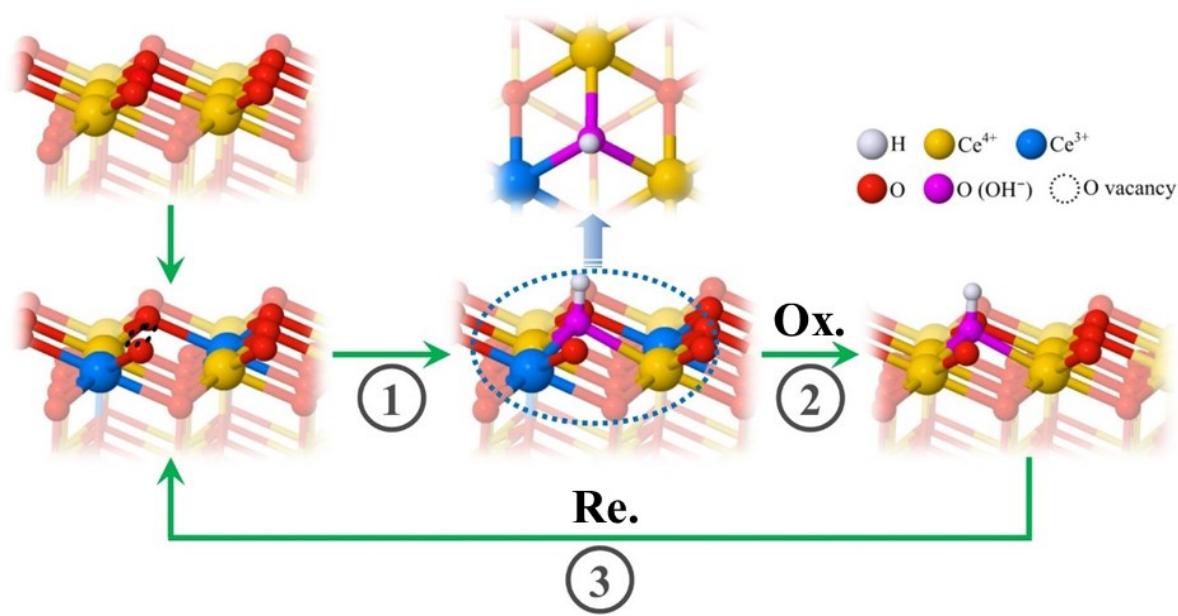

**Figure 7.** Schematic illustration showing changes in surface chemistry of $CeO_2$ driven by oxidation/reduction biomimetic reactions under neutral and basic conditions

However, in the case of acidic conditions (Figure 8), the reversibility of the $CeO_2$ surface is destroyed. Under acidic conditions, the surface $V_O^{\bullet\bullet}$ are filled only partially by $H_2O$ molecules which form one or two $Ce_{surface}$–$O_{water}$ bonds, rather than being filled completely by an $OH^-$ ion as under neutral and basic conditions. This means that, instead of annihilating the $V_O^{\bullet\bullet}$ as under neutral and basic conditions, the $CeO_2$ surface is not recovered to the non-defective-like structure and the $V_O^{\bullet\bullet}$ site still exists (Step 1). The $Ce^{3+}$ then are shifted back to $Ce^{4+}$ due to an oxidation biomimetic reaction (Step 2). A subsequent reduction biomimetic reaction will occur by shifting $Ce^{4+}$ to $Ce^{3+}$ and forming $V_O^{\bullet\bullet}$ (Step 3). However, since the $V_O^{\bullet\bullet}$ under acidic conditions are non-reversible and not able to be annihilated, the concentration of surface $V_O^{\bullet\bullet}$ will keep increasing over time, driven by reduction biomimetic reactions. Eventually, the concentration of surface $V_O^{\bullet\bullet}$ will reach the theoretical maximal concentration of intrinsic $V_O^{\bullet\bullet}$ (*viz.*, 25%[65]), and further formation of $V_O^{\bullet\bullet}$ will be hindered since the formation of a high concentration of $V_O^{\bullet\bullet}$ (*i.e.*, $V_O^{\bullet\bullet}$ clusters) is thermodynamically unfavorable[66]. As a result, the redox biomimetic reactions will be limited at acidic pH.

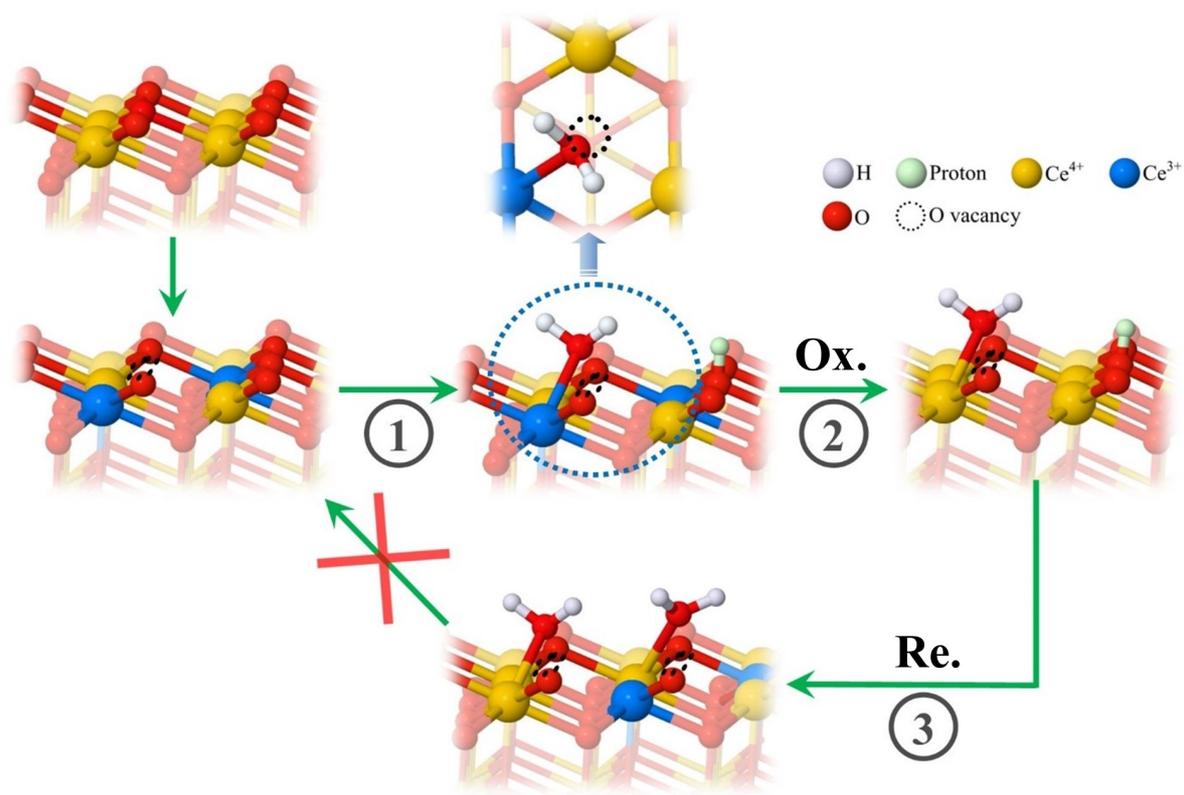

**Figure 8.** Schematic illustration showing changes in surface chemistry of $CeO_2$ driven by oxidation/reduction biomimetic reactions under acidic conditions

## Conclusions

In conclusion, pH has an important effect on the surface chemistry and redox biomimetic activities of $CeO_2$ owing to the distinct pH effect on the state of $V_O^{\bullet\bullet}$. Under acidic conditions, the $V_O^{\bullet\bullet}$ are filled only partially by the media $H_2O$ molecules, which destroys the reversibility of the surface structure, inhibiting the regenerative ability of $V_O^{\bullet\bullet}$. This limits the reversible cycling from $Ce^{4+}$ to $Ce^{3+}$ and destroys the redox biomimetic reaction cycle. In contrast, under neutral and basic conditions, the reversibility of the surface structure is preserved owing to the complete annihilation of $V_O^{\bullet\bullet}$. Consequently, the regenerative ability of $V_O^{\bullet\bullet}$ is retained, providing the prerequisite conditions for performing redox biomimetic reactions cyclically. This work thus provides a greater understanding of the pH influence on the surface chemistry of $CeO_2$ and creates an atomic view of the pH-induced multi-catalytic performance of $CeO_2$. It hence paves the way towards a better control and optimization of the material's performance in applications such as catalysis, diagnostics, and therapeutics, particularly cancer treatment.

## Supporting Information

Lattice parameters and optical indirect band gaps of $CeO_2$; formation energies for oxygen vacancies ($V_O^{\bullet\bullet}$) on $CeO_2$ {111} surface; total energies as a function of size of k-point mesh;

optimized geometries for one molecular layer and six molecular layers of $H_2O$ on $CeO_2$ {111} surface; initial geometries for locations of $Ce^{3+}$ relative to oxygen vacancies ($V_O^{\bullet\bullet}$); optimized geometries for media molecules on $CeO_2$ {111} surface; initial geometries for media molecules adsorbed on defective $CeO_2$ {111} surface.

**Author Contributions**

H. Ma carried out all the DFT calculations, data analysis, and wrote the paper. J. N. Hart, H. Ren, Z. Liu, P. Koshy, and C. C. Sorrell reviewed and revised the paper.

**Notes**

The authors declare that they have no conflict of interest.

**Acknowledgments**

This work was financially supported by the Australian Research Council Discovery Project scheme. This research was undertaken with the assistance of computational resources provided by the Australian Government through the National Computational Infrastructure (NCI) under the National Computational Merit Allocation Scheme.

**TOC Graphic**

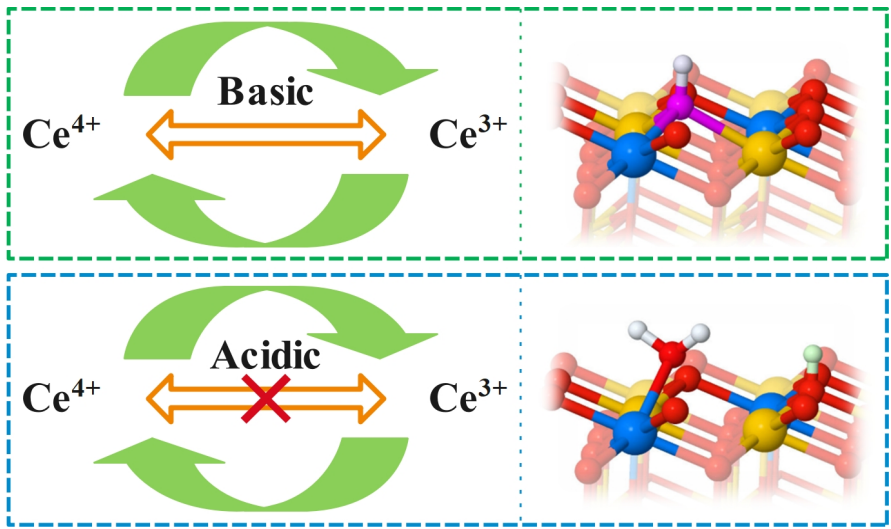